\documentclass[12pt]{article}

\usepackage{amssymb,amsmath}

\renewcommand{\appendix}{\setcounter{section}{0}
 \renewcommand{\thesection}{\Alph{section}} 
 \section*{Appendix} }

\renewcommand{\appendix}{\setcounter{section}{0}\renewcommand{\thesection}{}
\renewcommand{\thesubsection}{\Alph{subsection}.}
\section{Appendix}
}
\def\Appendix#1{
 			\setcounter{equation}{0}
 			\renewcommand{\theequation}{\thesubsection\arabic{equation}}
 			\subsection{#1}
 			}%
 			
\newcommand{\beq}[1]{  \\ {\tiny ({#1})}   \begin{equation} \label{#1} }
\newcommand{\beqa}[1]{ \\ {\tiny ({#1})}   \begin{eqnarray} \label{#1} }
\renewcommand{\beq}[1]{\begin{equation} \label{#1} }   
\renewcommand{\beqa}[1]{\begin{eqnarray} \label{#1} }  
\newcommand{\eeq}{\end{equation}}
\newcommand{\eeqa}{\end{eqnarray}}
\newcommand{\rf}[1]{(\ref{#1})}
\newcommand{\bosy}[1]{ \mbox{\boldmath ${#1}$} }

\def\doublespacing{\baselineskip=18pt} 
\usepackage{color}
\newif\ifpdf
\ifx\pdfoutput\undefined
      \pdffalse
\else
      \pdftrue
\fi
 
\ifpdf
    \pdfcatalog { /PageMode (/UseNone)
                  /OpenAction (fitbh)
    }
 
  \usepackage[pdftex]{graphicx}
  \usepackage[pdftex]{hyperref}
  \hypersetup{
   pdftitle={The closest anisotropic tensor},
    pdfsubject={anisotropic elastic tensors},
    pdfkeywords={elasticity tensors, Euclidean distance, Riemannian distance, 
    anisotropy},
    pdfauthor={Andrew N. Norris,
               <norris@rutgers.edu>},
    pdfpagemode={UseOutlines},
    bookmarksopen=true,
    colorlinks=true,
    urlcolor=rltblue,
    filecolor=rltgreen,
    linkcolor=rltred,
    citecolor=blue,
    pagecolor=red,
    urlcolor=cyan
  }
  \definecolor{rltred}{rgb}{0.75,0,0}
  \definecolor{rltgreen}{rgb}{0,0.5,0}
  \definecolor{rltblue}{rgb}{0,0,0.75}
\else
  \usepackage{graphicx}
\fi

\if@mathematic
   
\else
   
\fi

\begin{document}
\doublespacing 

\title{ 
{The isotropic material closest  to a \\ given anisotropic material} }

\author{Andrew N. Norris\\ \\    Mechanical and Aerospace Engineering, 
	Rutgers University, \\ Piscataway NJ 08854-8058, USA \,\, norris@rutgers.edu } 
	\date{}
\maketitle

\begin{abstract} 

The isotropic elastic moduli closest to a given anisotropic elasticity tensor are defined using three definitions of elastic distance, the standard Frobenius (Euclidean) norm,  the Riemannian distance for tensors,    and the log-Euclidean norm.  The closest moduli  are unique for the  Riemannian  and the log-Euclidean norms, independent of whether the difference in  stiffness or compliance is considered.  Explicit expressions for  the closest bulk and shear moduli are presented for cubic materials, and an algorithm is described for finding them for materials with arbitrary anisotropy.  The method is illustrated by 
application to  a  variety of materials, which are ranked according to their distance from isotropy. 

\end{abstract}

\section{Introduction}

The objective here is to  answer the question: what is the  isotropic material closest to a given anisotropic material.  In order to attempt an answer one needs a distance or length function which  measures  the difference between the elastic moduli of two materials.  The Euclidean norm provides a natural definition for distance, and using it one can find the elastic tensor of a given symmetry nearest to an anisotropic elastic tensor \cite{Gazis63}, see also \cite{Arts91,Helbig95,Cavallini99,Gangi00,Browaeys04}.   The Euclidean distance function is, however, not invariant under inversion, i.e. considering compliance instead of stiffness, and as such does not lead to a unique answer to the question posed.  To see this, let $\delta C_{ijkl}$ and $\delta S_{ijkl}$ be the elements of the fourth order tensors for the differences in elastic stiffness and compliance, respectively.  Define the length of a fourth order tensor with elements $T_{ijkl}$ by $(T_{ijkl} T_{ijkl})^{1/2}$. Then it is clear that the length using $\delta C_{ijkl}$ is not simply related to that of $\delta S_{ijkl}$.      

Recently and separately, Moakher \cite{Moakher04} and Arsigny et al. \cite{Arsigny:MICCAI:05} (see also \cite{Matthies95}) introduced two  distance functions for elasticity tensors which are unchanged whether one uses stiffness or compliance.  The two measures of elastic distances,  called the Riemannian distance \cite{Moakher04} and the log-Euclidean metric \cite{Arsigny:MICCAI:05},  
each  provide a means to unambiguously define the distance between any two elasticity tensors.  The focus here is on finding the  isotropic material closest to a given arbitrarily anisotropic material.  

The distance functions are  first reviewed 
 in Section \ref{sec1} along with the more common Frobenius or Euclidean norm.  The theory is developed in terms of matrices, with obvious application to tensors.   Preliminary results for elastic materials are presented in Section 3, where closed-form expressions are derived for the isotropic moduli closest to a given material of cubic symmetry.  The general problem for materials of arbitrary anisotropy is solved in Section 4, and applications to sample materials are described in Section 5.

\section{Matrix  distance functions}\label{sec1}

We begin with  ${\cal P}(n)$, the vector space of positive definite symmetric matrices in ${\mathbb M}^{n\times n}$, the space of $n\times n$ real matrices.  Recall that a matrix $\bf P$ is symmetric if ${\bosy{\it  x}}^T{\bf P}{\bosy{\it  y}}= {\bf y}^T{\bf P}{\bosy{\it  x}}$ for all ${\bosy{\it  x}}, {\bosy{\it  y}}$ in ${\mathbb R}^n$, and positive definite if ${\bosy{\it  x}}^T{\bf P}{\bosy{\it  x}}>0$ for all nonzero ${\bosy{\it  x}}\in {\mathbb R}^n$.  The spectral decomposition  is 
\beq{1}
{\bf P} = \sum\limits_{i=1}^n\, \lambda_i \, {\bf v}_i{\bf v}_i^T\, , 
\eeq
where $\lambda_i$ are the eigenvalues and ${\bf v}_i \in {\mathbb R}^n$ the eigenvectors, which satisfy  $\lambda_i > 0 $, ${\bf v}_i^T{\bf v}_j = \delta_{ij}$.   Functions of ${\bf P}$ can be readily found based on the diagonalized form, in particular, the logarithm of a matrix is  defined as 
\beq{2}
{\rm Log} \,{\bf P} = \sum\limits_{i=1}^n\, \ln \lambda_i \, {\bf v}_i{\bf v}_i^T\, . 
\eeq

Three distinct metrics for  positive definite symmetric matrices are  considered: the conventional Euclidean or Frobenius metric ${\rm d}_F$, the log-Euclidean norm ${\rm d}_L$ \cite{Arsigny:MICCAI:05} and the Riemannian distance ${\rm d}_R$  \cite{Moakher04}.  Thus, for any pair ${\bf A},{\bf B} \in {\cal P}(n)$ 
\begin{subequations}\label{3}
\beqa{3a}
{\rm d}_F ({\bf A},{\bf B}) &=& \left\| {\bf A} -{\bf B} \right\|,
\\
{\rm d}_L ({\bf A},{\bf B}) &=& \left\| {\rm Log}({\bf A}) - {\rm Log}({\bf B}) \right\|,  
\label{3L}
\\
{\rm d}_R ({\bf A},{\bf B}) &=& \left\| {\rm Log}({\bf A}^{-1/2}{\bf B}{\bf A}^{-1/2}) \right\|,  
\label{3b}
\eeqa
\end{subequations}
where the norm  is defined  $\left\| {\bf M} \right\|= [{\rm tr}({\bf M}^T{\bf M})]^{1/2}$ for any 
 ${\bf M}\in {\mathbb M}^{n\times n}$. 
 The distance function ${\rm d}_R$ is a consequence of  the scalar product  
 \beq{5071}
 \langle {\bf M}_1,\, {\bf M}_2\rangle_{\bf P} \equiv  
 {\rm tr}({\bf P}^{-1}{\bf M}_1{\bf P}^{-1}{\bf M}_2),
 \eeq
  for ${\bf P}\in {\cal P}(n)$ and symmetric 
 ${\bf M}_1,\, {\bf M}_2 \in {\mathbb M}^{n\times n}$,  
 and is also related to the exponential map \cite{lang98,Moakher04}. The metric  ${\rm d}_L$  is associated with the Lie group on ${\cal P}(n)$ defined by the following multiplication that preserves symmetry and positive definiteness \cite{Arsigny:MICCAI:05} :
\beq{5081}
{\bf P}_1 \odot {\bf P}_2 \equiv \exp \left( {\rm Log}({\bf P}_1) + {\rm Log}({\bf P}_2)\right),
\quad 
{\bf P}_1,\, {\bf P}_2\, 
 \in \, {\cal P}(n) . 
 \eeq

The three distance functions possess the  properties expected of a distance function ${\rm d}$: (i) it is symmetric with respect to its arguments, ${\rm d} ({\bf A},{\bf B})= {\rm d} ({\bf B},{\bf A})$; (ii) non-negative ${\rm d} ({\bf A},{\bf B}) \ge 0$ with equality iff ${\bf A}={\bf B}$; 
(iii) it is invariant under a change of basis, 
${\rm d} ({\bf Q}{\bf A}{\bf Q}^T,{\bf Q}{\bf B}{\bf Q}^T)  = {\rm d} ({\bf A},{\bf B})$ for all 
orthogonal ${\bf Q} \in {\mathbb M}^{n\times n}$,  ${\bf Q}{\bf Q}^T = {\bf Q}^T{\bf Q}
= {\bf I}$; and (iv) it satisfies the triangle inequality  ${\rm d} ({\bf A},{\bf C}) \le {\rm d} ({\bf A},{\bf B}) + {\rm d} ({\bf B},{\bf C})$ for all ${\bf A},{\bf B}, {\bf C}\in {\cal P}(n)$.

The Riemannian and log-Euclidean distances have additional properties not shared with ${\rm d}_F$: 
\begin{subequations}\label{6}
\beqa{6a}
   {\rm d}_{L,R} (a{\bf A},a{\bf B})  &=& {\rm d}_{L,R} ({\bf A},{\bf B})\, , \quad a\in \mathbb{R}_+, 
\\  
{\rm d}_{L,R} ({\bf A}^b,{\bf B}^b) &=& |b|\, {\rm d}_{L,R} ({\bf A},{\bf B}) \,,\quad b\in \mathbb{R}, 
\label{6b}
\eeqa
\end{subequations}
where $ {\rm d}_{L,R}$ signifies either $ {\rm d}_{L}$ or $ {\rm d}_{R}$. 
Thus ${\rm d}_{L}$ and ${\rm d}_{R}$ are bi-invariant metrics, i.e. distances  are invariant under 
multiplication and inversion.   This  property  makes them consistent and unambiguous  metrics for elasticity tensors. 
Moakher \cite{Moakher04} introduced another bi-invariant distance function, the 
 Kullback-Leibler metric, but it does not satisfy the triangle inequality, and we do not consider it here.  
 
The distance function ${\rm d}_R$  can be expressed in  alternative forms by using  the property   ${\bf B} ({\rm Log} \,{\bf A}){\bf B}^{-1}= {\rm Log} \,({\bf B}{\bf A} {\bf B}^{-1})$, e.g., 
\beq{836}
{\rm d}_R ({\bf A},{\bf B}) = \big[{\rm tr} \, {\rm Log}^2({\bf A}^{-1}{\bf B})\big]^{1/2}
= \big[{\rm tr} \, {\rm Log}^2({\bf B}^{-1}{\bf A})\big]^{1/2}, 
\eeq
or in terms of  eigenvalues, using eqs. \rf{2} and   \rf{3b}, 
\beq{4}
{\rm d}_R ({\bf A},{\bf B}) = \bigg[ \sum\limits_{i=1}^n\, (\ln \lambda_i)^2\bigg]^{1/2}, 
\eeq
where $\lambda_i$, $i=1,2,\ldots, n$ are the eigenvalues of 
${\bf P} = {\bf A}^{-1/2}{\bf B}{\bf A}^{-1/2}$, or equivalently, of the matrices ${\bf A}^{-1}{\bf B}$, 
${\bf B}^{-1}{\bf A}$, ${\bf A}{\bf B}^{-1}$, etc. 
Note that ${\rm d}_{R}$ also satisfies 
\beq{834}
{\rm d}_R ({\bf S}{\bf A}{\bf S}^T,{\bf S}{\bf B}{\bf S}^T) =  {\rm d}_R ({\bf A},{\bf B}) \, ,\quad \forall \, \, {\rm invertible }\, \, {\bf S} \in {\mathbb M}^{n\times n}.
\eeq

\section{Preliminary examples}

The remainder of the paper is concerned with applications to elasticity, with $n=6$. 

\subsection {Definition of elastic moduli}

$6\times 6$ symmetric matrices are used to describe  elastic moduli, whether of stiffness or compliance.  The matrix representation is based on Kelvin's \cite{kelvin} observation in 1856 that the twenty one coefficients of the elasticity define a quadratic form (the energy) in the six strains, and therefore possess six ``principal strains" \cite{kelvin}.    Although Kelvin did not write the elasticity tensor explicitly as a symmetric positive definite matrix,  the idea has proven  useful and has been developed extensively, notably by Rychlewski \cite{ry} and  Mehrabadi and Cowin \cite{c3}.   The notation of Mehrabadi and Cowin \cite{c3} is employed here.  Thus, the matrix ${\widehat{\bf C}}\in {\cal P}(6)$ represents the elastic stiffness, and its inverse is the elastic compliance, $\widehat{\bf S}$, satisfying 
\beq{sc}
\widehat{\bf S}{\widehat{\bf C}} = {\widehat{\bf C}} \widehat{\bf S} = \widehat{\bf I}, 
\qquad {\rm where}\,\, \widehat{\bf I}  = {\rm diag}(1,1,1,1,1,1).
\eeq
 The elements of the elastic stiffness matrix are 
\beq{a1}
{\widehat{\bf C}}  = 
\begin{pmatrix}
\hat{c}_{11} & \hat{c}_{12} & \hat{c}_{13} & 
 \hat{c}_{14} & \hat{c}_{15} &  \hat{c}_{16} 
\\ & & & & & \\
\hat{c}_{12} & \hat{c}_{22} & \hat{c}_{23} & 
  \hat{c}_{24} &  \hat{c}_{25} &  \hat{c}_{26} 
\\ & & & & & \\
\hat{c}_{13} & \hat{c}_{23} & \hat{c}_{33} & 
  \hat{c}_{34} &  \hat{c}_{35} &  \hat{c}_{36} 
\\ & & & & & \\
  \hat{c}_{14} &  \hat{c}_{24} &  \hat{c}_{34} 
 & \hat{c}_{44} & \hat{c}_{45} & \hat{c}_{46}
\\ & & & & & \\
  \hat{c}_{15} &  \hat{c}_{25} &  \hat{c}_{35} 
 & \hat{c}_{45} & \hat{c}_{55} & \hat{c}_{56}
\\ & & & & & \\
  \hat{c}_{16} &  \hat{c}_{26} &  \hat{c}_{36} 
 & \hat{c}_{46} & \hat{c}_{56} & \hat{c}_{66}
\end{pmatrix} 
=\begin{pmatrix}
c_{11} & c_{12} & c_{13} & 
 2^{\frac12} c_{14} & 2^{\frac12} c_{15} & 2^{\frac12} c_{16} 
\\ & & & & & \\
c_{12} & c_{22} & c_{23} & 
 2^{\frac12} c_{24} & 2^{\frac12} c_{25} & 2^{\frac12} c_{26} 
\\ & & & & & \\
c_{13} & c_{23} & c_{33} & 
 2^{\frac12} c_{34} & 2^{\frac12} c_{35} & 2^{\frac12} c_{36} 
\\ & & & & & \\
 2^{\frac12} c_{14} & 2^{\frac12} c_{24} & 2^{\frac12} c_{34} 
 & 2c_{44} & 2c_{45} & 2c_{46}
\\ & & & & & \\
 2^{\frac12} c_{15} & 2^{\frac12} c_{25} & 2^{\frac12} c_{35} 
 & 2c_{45} & 2c_{55} & 2c_{56}
\\ & & & & & \\
 2^{\frac12} c_{16} & 2^{\frac12} c_{26} & 2^{\frac12} c_{36} 
 & 2c_{46} & 2c_{56} & 2c_{66}
\end{pmatrix}  , 
\eeq
where $c_{ij}, \ i,j=1,2,\ldots 6$ are the coefficients  in the Voigt notation.

Before considering materials of arbitrary anisotropy it is instructive to examine the distance functions for materials possessing the simplest type of anisotropy: cubic symmetry.  
Materials of cubic symmetry are described by  three independent moduli: 
$hat{c}_{11}=hat{c}_{22}=hat{c}_{33}$, $hat{c}_{12}=hat{c}_{23}=hat{c}_{13}$, 
$hat{c}_{44}=hat{c}_{55}=hat{c}_{66}$, with the rest equal to zero. 
The three moduli commonly used are the bulk modulus $\kappa$ and the two distinct  shear moduli  $\mu$ and $\eta$, which are related to the matrix elements by 
\beq{iso}
3\kappa = \hat{c}_{11}+2 \hat{c}_{12}, \qquad 2\mu = \hat{c}_{44},\qquad 2\eta = \hat{c}_{11}- \hat{c}_{12}.  
\eeq
Isotropic materials have only two independent moduli $\kappa$, $\mu$, and are of the same form as for cubic materials  with the restriction $ \hat{c}_{11}- \hat{c}_{12}-\hat{c}_{44}=0$, or equivalently, $\eta=\mu$.

A concise notation is used for isotropic and cubic matrices, based upon Walpole's \cite{walpole84} general scheme for performing algebra with elasticity tensors.  Define 
the  matrices $\widehat{\bf J} $, $\widehat{\bf K} $, $\widehat{\bf L} $ and $ \widehat{\bf M}$ by 
\begin{subequations}\label{911}
\beqa{911a}
&&\widehat{\bf K} = \widehat{\bf I}-\widehat{\bf J}, 
\qquad \widehat{\bf J} = {\bf u}{\bf u}^T, \qquad  {\rm where}\,\,  
{\bf u} = (\tfrac{1}{\sqrt{3}}, \,\tfrac{1}{\sqrt{3}},\, \tfrac{1}{\sqrt{3}}, \, 0,\, 0,\, 0)^T,
\\
&&\widehat{\bf M} = \widehat{\bf K}-\widehat{\bf L},\qquad  \,\ \widehat{\bf L} = {\rm diag}\,(0,\, 0,\, 0,\, 1,\, 1,\, 1)\, .  \label{911b}
\eeqa
\end{subequations}
Note that $\widehat{\bf I} $ and $ \widehat{\bf J}$ correspond, respectively, to  
the  fourth order isotropic symmetric tensors with components
$I_{ijkl} = (\delta_{ik}\delta_{jl}+\delta_{il}\delta_{jk})/2$ and 
$J_{ijkl} = (1/3)\delta_{ij}\delta_{kl}$.  
Elastic moduli of isotropic and cubic  materials are of the generic form 
\begin{subequations}\label{111}
\beqa{111a} 
{\widehat{\bf C}}_{\rm iso} (3\kappa,\, 2\mu)
&\equiv &3\kappa \, \widehat{\bf J} + 2\mu \, \widehat{\bf K}, \qquad\qquad\quad \, \kappa,\, \mu  >0 
\, , 
\\
 {\widehat{\bf C}}_{\rm cub} (3\kappa,\, 2\mu,\, 2\eta)
&\equiv &3\kappa \, \widehat{\bf J} + 2\mu \, \widehat{\bf L}+ 2\eta \, \widehat{\bf M}, \qquad \kappa,\, \mu, \, \eta  >0  \, . 
\label{111b}
\eeqa
\end{subequations}
The isotropic matrices $\{ \widehat{\bf J}, \widehat{\bf K}\}$ are idempotent and their  matrix product vanishes,  ${\widehat{\bf J}}^2 =\widehat{\bf J}$, ${\widehat{\bf K}}^2 =\widehat{\bf K}$, $\widehat{\bf J} \widehat{\bf K}= \widehat{\bf K} \widehat{\bf J}= 0$.  Similarly,  it may be checked that 
the three basis matrices for cubic materials $\{ \widehat{\bf J},  \widehat{\bf L}, \widehat{\bf M}\}$ are idempotent and have zero mutual products. 
The algebra of matrix multiplication for isotropic and cubic materials follows from these basic multiplication tables: 
\beqa{806}
&\begin{matrix}
 \\
 \hline
 \widehat{\bf J}
 \\
 \widehat{\bf K}
\end{matrix}
\left|
\begin{matrix}
\widehat{\bf J}
&
 \widehat{\bf K}
 \\
 \hline
 \widehat{\bf J} & 0 
 \\
 0 & \widehat{\bf K}  
  \end{matrix}
  \right.
  \qquad \qquad \qquad
  \begin{matrix}
 \\
 \hline
 \widehat{\bf J}
 \\
 \widehat{\bf L}
 \\
 \widehat{\bf M}
 \end{matrix}
\left|
 \begin{matrix}
\widehat{\bf J}
&
 \widehat{\bf L}
&
 \widehat{\bf M}
 \\
 \hline
 \widehat{\bf J} & 0 & 0 
 \\
 0 & \widehat{\bf L} &  0 
 \\
  0 & 0 & \widehat{\bf M} 
  \end{matrix} \, . 
  \right.&
\nonumber 
\eeqa
  Thus, inverses are  $\widehat{\bf S}_{\rm cub} = {\widehat{\bf C}}_{\rm cub}^{-1} = \widehat{\bf C}_{\rm cub} (\frac1{3\kappa},\, \frac1{2\mu},\, \frac1{2\eta})$, 
$\widehat{\bf S}_{\rm iso} =  \widehat{\bf C}_{\rm iso} (\frac1{3\kappa},\, \frac1{2\mu})$, and products are 
\begin{subequations}\label{112}
\beqa{112a} 
{\widehat{\bf C}}_{\rm iso}^{-1} (3\kappa_1,\, 2\mu_1)
{\widehat{\bf C}}_{\rm iso} (3\kappa_2,\, 2\mu_2)
&\equiv &\frac{\kappa_2}{\kappa_1} \, \widehat{\bf J} + \frac{\mu_2}{\mu_1} \, \widehat{\bf K} \, , 
\\
 {\widehat{\bf C}}_{\rm cub}^{-1} (3\kappa_1,\, 2\mu_1,\, 2\eta_1) 
{\widehat{\bf C}}_{\rm cub} (3\kappa_2,\, 2\mu_2,\, 2\eta_2)
&\equiv &\frac{\kappa_2}{\kappa_1} \, \widehat{\bf J} + \frac{\mu_2}{\mu_1} \, \widehat{\bf L}+ \frac{\eta_2}{\eta_1} \, \widehat{\bf M}   \, . 
\label{112b}
\eeqa
\end{subequations}
Results for isotropic materials follow from those for cubic with $\eta = \mu$.  For the sake of simplicity and brevity we therefore focus on properties  for cubic materials in the next subsection. 

\subsection{Elastic distance for cubic and isotropic materials}

Consider two  cubic materials with moduli 
$\widehat{\bf C}_1 = {\widehat{\bf C}}_{\rm cub} (3\kappa_1, 2\mu_1, 2\eta_1)$
and  $\widehat{\bf C}_2$ $= {\widehat{\bf C}}_{\rm cub} $ $(3\kappa_2, $ $2\mu_2, $ $2\eta_2)$. 
The Euclidean distance function  of eq. \rf{3a} follows from the above properties and the relations tr$\widehat{\bf J}=1$,  tr$\widehat{\bf L}=3$, tr$\widehat{\bf M}=2$.  Similarly,  the Riemannian and log-Euclidean distances  follows from the identities 
\beq{113} 
{\rm Log}({\widehat{\bf C}}_2  ) - {\rm Log}({\widehat{\bf C}}_1  )  = 
{\rm Log}\, {\widehat{\bf C}}_1^{-1}  
{\widehat{\bf C}}_2  
=\ln \big(\frac{\kappa_2}{\kappa_1}\big) \, \widehat{\bf J} + \ln \big(\frac{\mu_2}{\mu_1}\big) \, \widehat{\bf L}+ \ln \big(\frac{\eta_2}{\eta_1}\big) \, \widehat{\bf M}   \, .
\eeq
Thus, the distances functions  are
\begin{subequations}\label{22}
\beqa{22a}
{\rm d}_F (\widehat{\bf C}_1,\, \widehat{\bf C}_2)
&=& \big[  (3\kappa_1-3\kappa_2)^2 + 3(2\mu_1-2\mu_2)^2+ 2(2\eta_1-2\eta_2)^2  \big]^{1/2}, 
\quad
\\ 
{\rm d}_{L,R} (\widehat{\bf C}_1,\, \widehat{\bf C}_2)
&=& \big[  (\ln \frac{\kappa_2}{\kappa_1})^2 + 3(\ln \frac{\mu_2}{\mu_1})^2+ 2(\ln \frac{\eta_2}{\eta_1})^2  \big]^{1/2}\, . 
\label{22b}
\eeqa
\end{subequations}
It is clear  that ${\rm d}_L $  and ${\rm d}_R $ are invariant under inversion, 
${\rm d}_{L,R} (\widehat{\bf S}_1,\, \widehat{\bf S}_2) = {\rm d}_{L,R} (\widehat{\bf C}_1,\, \widehat{\bf C}_2)$.  
Note that the first identity in \rf{113} is a consequence of the fact that ${\widehat{\bf C}}_1$ and ${\widehat{\bf C}}_2$ commute, which is not true in general for material symmetries lower than cubic.  

What is the isotropic material closest to a given cubic material?  The answer may be found by considering  the distance functions between  an arbitrary cubic stiffness  
${\widehat{\bf C}}_{\rm cub} (3\kappa, 2\mu, 2\eta)$ and the isotropic stiffness 
${\widehat{\bf C}}_{\rm iso} (3\kappa_*, 2\mu_*)$.  The same question will  also be considered for the compliances. 
Minimizing with respect to the isotropic moduli 
$\kappa_*$, $\mu_*$ yields
\begin{subequations}\label{24}
\beqa{24a}
&&\min_{\kappa_*,\, \mu_*}\, {\rm d}_{L,R}\bigg( \widehat{\bf C}_{\rm cub} ,\, \widehat{\bf C}_{\rm iso} (3\kappa_*,2\mu_*)\bigg) 
= \min_{\kappa_*,\, \mu_*}\, {\rm d}_{L,R}\bigg( \widehat{\bf S}_{\rm cub} ,\, \widehat{\bf S}_{\rm iso} \bigg) 
= \sqrt{\tfrac65}\, \left|\ln \frac{\mu}{\eta}\right |,
\\
&&\min_{\kappa_*,\, \mu_*}\, {\rm d}_F\bigg( \widehat{\bf C}_{\rm cub} ,\, \widehat{\bf C}_{\rm iso} (3\kappa_*,2\mu_*)\bigg) 
=  \sqrt{\tfrac65}\, \left|2\mu -2\eta\right|,
\\
&&\min_{\kappa_*,\, \mu_*}\, {\rm d}_F\bigg( \widehat{\bf C}_{\rm cub}^{-1} ,\, \widehat{\bf C}_{\rm iso}^{-1} (3\kappa_*,2\mu_*)\bigg) 
=  \sqrt{\tfrac65}\, \left|\frac1{2\mu} -\frac1{2\eta}\right|\, . 
\label{24b}
\eeqa
\end{subequations}
Denote the values of the closest isotropic moduli by  $(\kappa_L, \mu_L )$, $(\kappa_R, \mu_R)$ for ${\rm d}_L$, ${\rm d}_R$, and   
$(\kappa_A, \mu_A)$ or $(\kappa_H, \mu_H)$ for ${\rm d}_F$ depending as the stiffness $(A)$ or its inverse $(H)$ is used.  Thus, 
\beq{25}
\kappa_{L,R,A,H}=\kappa,\quad  \mu_{L,R} = (\mu^3  \eta^2)^{1/5}  ,
\quad  \mu_A = \frac35 \mu + \frac25 \eta ,
\quad  \frac1{\mu_H} = \frac3{5 \mu} + \frac2{5 \eta}\, .  
\eeq
Equations \rf{24} and \rf{25} show clearly that  the ``closest" isotropic material using the Frobenius metric is ambiguous because it depends on whether one uses  stiffness or compliance. Each gives a different isotropic material since $\mu_H < \mu_{L,R} < \mu_A$ for $\mu -\eta \ne 0$.   The Riemannian and log-Euclidean metrics gives the same unique ``closest" isotropic material, regardless of whether  the  stiffness or  the compliance are used.  The fact that they agree is particular to the case of cubic symmetry, as noted above, and is not true in general.

In summary, the closest isotropic material 
to a given cubic material,  in the sense of ${\rm d}_R$ and ${\rm d}_L$,  is defined by moduli $\kappa_R= \kappa_L = \frac13 (\hat{c}_{11}+2 \hat{c}_{12})$ and $\mu_R= \mu_L = \frac12 \big[ \hat{c}_{44}^3\, (\hat{c}_{11}- \hat{c}_{12})^2 \big]^{1/5}$,   
and the distance from isotropy is 
${\rm d}_{L,R} =  \sqrt{6/5}\, \left|\ln \frac{\hat{c}_{11}- \hat{c}_{12}}{\hat{c}_{44}}\right|$. 
These results will be  generalized to materials of arbitrary anisotropy next. 

\section{Closest isotropic moduli}

We now turn to the more general question of finding the isotropic material closest to a given anisotropic material characterized by $\widehat{\bf C}$ or its inverse $\widehat{\bf S}$.      The solution using the  Euclidean metric  is relatively simple, and is considered first. 

\subsection{Minimum Frobenius   distances}

The closest isotropic elastic moduli are assumed to be of general isotropic  form 
${\widehat{\bf C}}_{\rm iso} (3\kappa ,\, 2\mu)$, see eq. \rf{111}.  
The bulk and shear moduli are found by minimizing 
${\rm d}_F( {\widehat{\bf C}}_{\rm iso},\, {\widehat{\bf C}})$, which implies 
\beq{62}
  3\kappa \, {\rm tr} \, \widehat{\bf J} = {\rm tr} \, \widehat{\bf J}{\widehat{\bf C}} , 
  \qquad
  2\mu \, {\rm tr} \, \widehat{\bf K} = {\rm tr} \, \widehat{\bf K}{\widehat{\bf C}} \, . 
\eeq
Using suffix $A$ to indicate that the minimization is in the arithmetic sense (in line with  \cite{Moakher04}), 
\begin{subequations}\label{63}
\beqa{63a}
  9\kappa_A    &=& \hat{c}_{11}+\hat{c}_{22}+\hat{c}_{33}+2(\hat{c}_{23}+\hat{c}_{31}+\hat{c}_{12})\, ,
  \\
  30\mu_A   &=&  2(
  \hat{c}_{11}+\hat{c}_{22}+\hat{c}_{33}-\hat{c}_{23}-\hat{c}_{31}-\hat{c}_{12})
  +3( \hat{c}_{44} +  \hat{c}_{55} +  \hat{c}_{66} )
  \,,  \quad
  \label{63b}
\eeqa
\end{subequations}
which  are well known, e.g. \cite{fed}.   Similarly, the closest isotropic elastic compliance can be determined by minimizing ${\rm d}_F( {\widehat{\bf C}}_{\rm iso}^{-1},\, {\widehat{\bf C}}^{-1})$.  Denoting the isotropic moduli with the suffix $H$ for harmonic,  
\begin{subequations}\label{64}
\beqa{64a}
  1/{\kappa_H}    &=& \hat{s}_{11}+\hat{s}_{22}+\hat{s}_{33}+2(\hat{s}_{23}+\hat{s}_{31}+\hat{s}_{12})\, ,
  \\
  15/(2\mu_H)  &=&  2(
  \hat{s}_{11}+\hat{s}_{22}+\hat{s}_{33}-\hat{s}_{23}-\hat{s}_{31}-\hat{s}_{12})
  +3( \hat{s}_{44} +  \hat{s}_{55} +  \hat{s}_{66} )
  \, .  \quad
  \label{64b}
\eeqa
\end{subequations}
The Euclidean distance does not provide a unique closest isotropic material, although the  values in  eqs. \rf{63} and \rf{64} are sometimes considered as bounds.  Equations \rf{62} and \rf{63} also agree with the special case discussed above for cubic materials, eq. \rf{25}. 

\subsection{Minimum log-Euclidean   distance}

The isotropic elasticity ${\widehat{\bf C}}_{\rm iso} (3\kappa_L ,\, 2\mu_L)$ is found using the same methods as above by  
replacing $ {\widehat{\bf C}}_{\rm iso}$ and $ {\widehat{\bf C}}$ with 
$ {\rm Log} ({\widehat{\bf C}}_{\rm iso})$ and $ {\rm Log}({\widehat{\bf C}})$, respectively. Thus, 
\beq{623}
  \log (3\kappa_L)  = {\rm tr} \, \widehat{\bf J}{\rm Log} ({\widehat{\bf C}}) , 
  \qquad
  5\, \log (2\mu_L) = {\rm tr} \, \widehat{\bf K}{\rm Log} ({\widehat{\bf C}}) \, . 
\eeq
Adding the two equations and using $\widehat{\bf J} + \widehat{\bf K} = \widehat{\bf I}$, implies the identity
\beq{624}
{\rm det}({\widehat{\bf C}}_{\rm iso} ) = {\rm det}({\widehat{\bf C}})  . 
\eeq
Thus, we have  explicit formulae for the closest moduli,
\beq{625}
 \kappa_L  = \tfrac13\, \exp \left( {\rm tr} \, \widehat{\bf J}{\rm Log} ({\widehat{\bf C}}) \right), 
  \qquad
\mu_L = \tfrac12\, \exp \left( \tfrac15\, {\rm tr} \, \widehat{\bf K}{\rm Log} ({\widehat{\bf C}}) \right)\, . 
\eeq

\subsection{The minimum Riemannian distance}

We look for moduli of the form 
\beq{501}
 {\widehat{\bf C}}_{\rm iso} (3\kappa_R ,\, 2\mu_R) 
= 3\kappa_R \, \widehat{\bf J} + 2\mu_R\, \widehat{\bf K} \, ,  
\eeq
which minimize ${\rm d}_R( {\widehat{\bf C}}_{\rm iso},\, {\widehat{\bf C}})$.  Differentiating the expression 
\beq{502}
{\rm d}_R^2 ( {\widehat{\bf C}}_{\rm iso},\, {\widehat{\bf C}})
= {\rm tr}\, \big[ {\rm Log}({\widehat{\bf C}}_{\rm iso}^{-1}{\widehat{\bf C}})\, 
{\rm Log}({\widehat{\bf C}}_{\rm iso}^{-1}{\widehat{\bf C}}) \big] \, , 
\eeq
with respect to $\kappa_R$ and $\mu_R$ separately, and using 
${\rm Log}({\bf A})' = {\bf A}^{-1} {\bf A}'$ for ${\bf A} \in {\cal P}(n)$,  implies respectively
\beq{5051}
{\rm tr}\, \big[ {\widehat{\bf C}}_{\rm iso}^{-1} \widehat{\bf J}\, 
{\rm Log}({\widehat{\bf C}}_{\rm iso}^{-1}{\widehat{\bf C}}) \big] = 0\, , 
\qquad 
{\rm tr}\, \big[ {\widehat{\bf C}}_{\rm iso}^{-1}\widehat{\bf K}\, 
{\rm Log}({\widehat{\bf C}}_{\rm iso}^{-1}{\widehat{\bf C}}) \big] = 0\, .  
\eeq
Further simplification yields 
\beq{505}
{\rm tr}\, \big[ \widehat{\bf J}\, 
{\rm Log}({\widehat{\bf C}}_{\rm iso}^{-1}{\widehat{\bf C}}) \big] = 0\, , 
\qquad
{\rm tr}\, \big[ \widehat{\bf K}\, 
{\rm Log}({\widehat{\bf C}}_{\rm iso}^{-1}{\widehat{\bf C}}) \big] = 0\, .  
\eeq
These conditions, which  are necessary for a minimum,   can be simplified as follows.  Define the eigenvalues and associated eigenvectors by the diagonalization
\beq{507}
{\widehat{\bf C}}_{\rm iso}^{-1/2}{\widehat{\bf C}} {\widehat{\bf C}}_{\rm iso}^{-1/2}= \sum\limits_{i=1}^n\, \lambda_i \, {\bf v}_i{\bf v}_i^T\, . 
\eeq
Adding the two conditions \rf{505}  using the identity $\widehat{\bf I}= \widehat{\bf J}+\widehat{\bf K}$, along with the expression  \rf{2} for the logarithm of a matrix, yields 
\beq{508}
\prod\limits_{i=1}^n\,   \lambda_i = 1\, . 
\eeq
A second condition follows by direct substitution from \rf{507} into $\rf{505}_1$, giving 
 \beq{509}
\prod\limits_{i=1}^n\,   \lambda_i^{\alpha_i} = 1\, ,  \qquad 
\alpha_i \equiv {\bf v}_i^T{\widehat{\bf J}}{\bf v}_i,\,\, \, i=1,2,\ldots n. 
\eeq
Note that $0\le \alpha_i \le 1$ and $\alpha_i $ form a partition of unity, 
\beq{510}
\sum\limits_{i=1}^n\,  \alpha_i =1\, .  
\eeq
This follows from the representation 
$\widehat{\bf J} = {\bf u}{\bf u}^T$ where the unit $6-$vector ${\bf u}$ is defined in eq. \rf{911a}.  
Thus, the minimal isotropic moduli are found by satisfying the two simultaneous equations \rf{508} and \rf{509}.  
We now show how the first of these two conditions can be met, leaving one condition to satisfy.  

Let 
\beq{211}
{\widehat{\bf C}}_{\rm iso} = 3\kappa_R \,\big(      \widehat{\bf J} + \rho^{-2} \widehat{\bf K} \big)\, ,  
\eeq
where $\rho \ge 0$ is defined by 
\beq{2112}
\rho^2 =  \frac {3\kappa_R}{2\mu_R}    =  \frac {1+\nu_R}{1-2\nu_R} \, , 
\eeq
and $\nu_R$ is the Poisson's ratio of the minimizer. 
The reason  this form  for ${\widehat{\bf C}}_{\rm iso}$ is chosen is so that 
${\widehat{\bf C}}_{\rm iso}^{-1/2} =  (3\kappa_R)^{-1/2} \,\big(   \widehat{\bf J} + \rho \widehat{\bf K} \big)$.  Hence, 
 the eigenvalues of \rf{507} are of the form
\beq{212}
\lambda_i =   \frac{\bar{\lambda}_i (\rho)}{3\kappa_R}\, ,  
\eeq
where the normalized eigenvectors $\bar{\lambda}_i=\bar{\lambda}_i (\rho)$, and the (unchanged)  eigenvectors ${\bf v}_i$, $i=1,2,\ldots, n=6$ are defined by   
\beq{215}
3\kappa_R\, {\widehat{\bf C}}_{\rm iso}^{-1/2} \widehat{\bf C}{\widehat{\bf C}}_{\rm iso}^{-1/2} =  \big(   \widehat{\bf J} +  \rho \widehat{\bf K}\big)\, {\widehat{\bf C}} \,\big(   \widehat{\bf J} +  \rho \widehat{\bf K}\big)
=   \sum\limits_{i=1}^n\, \bar{\lambda}_i \, {\bf v}_i{\bf v}_i^T\, .\  
\eeq

Turning to the first condition, \rf{508},  it is automatically satisfied if the bulk modulus is given by 
\beq{214}
3\kappa_R = 
\bigg( \prod\limits_{i=1}^n\, \bar{\lambda}_i \bigg)^{1/n}\, .  
\eeq
It remains to determine  $\rho$ from the second stationary condition, eq. \rf{509}, which can be expressed in terms of the modified eigenvalues as 
\beq{511}
\prod\limits_{i=1}^n\,  \bar{\lambda}_i^{(\alpha_i - 1/n)} = 1\, . 
\eeq
Equation \rf{511}   involves the eigenvectors ${\bf v}$ through the inner products $\alpha_i$.  However,  $\alpha_i$ vanishes identically for eigenvectors of  {\it deviatoric} form - in fact the definition of a deviatoric eigenvector is  $\alpha_i =0$ \cite{c3}.  Conversely, $\alpha_i = 1$  for purely {\it dilatational} eigenvectors \cite{c3}, i.e., eigenvectors parallel to $\bf u$ of eq. \rf{911a}. 

The solution to eq. \rf{511} may be found numerically by searching for the zero  over the permissible range for the Poisson's ratio: $-1 < \nu_R < 1/2$.  The minimizing moduli $\kappa_R$ and $\mu_R$ then follow from   eqs. \rf{214} and \rf{2112}, or more directly, 
\beq{5111}
3\kappa_R = \rho^{5/3}\, \big( {\rm det}\, \widehat{\bf C}\big)^{1/6},
\qquad
2\mu_R = \rho^{-1/3}\, \big( {\rm det}\, \widehat{\bf C}\big)^{1/6}\, , 
\eeq
and the minimal  distance between ${\widehat{\bf C}}_{\rm iso}$ and ${\widehat{\bf C}}$ is given by 
\beq{213}
{\rm d}_R ({\widehat{\bf C}}_{\rm iso},\, {\widehat{\bf C}}) = \frac1{n} \, \bigg[ \sum\limits_{i=1}^n\,  \ln^2 \bigg(
(\bar{\lambda}_i)^{-n} {\prod\limits_{j=1}^n\, \bar{\lambda}_j }\bigg)  \bigg]^{1/2} \, \qquad (n=6).   
\eeq

We next demonstrate the application of the above procedure to the case of a given elasticity matrix of cubic symmetry.  

\subsection{Example: cubic materials}

Substituting the assumed form ${\widehat{\bf C}}= {\widehat{\bf C}}_{\rm cub}$ from eq. \rf{111b} into the explicit formulae of eq. \rf{625} for the closest moduli in the  log-Euclidean sense, it is a straightforward matter to show that these reproduce the results determined directly, in eq. \rf{25}.
Regarding the closest moduli using the Riemannian distance, 
the matrix in eq. \rf{215} follows by using the algebra for cubic matrices, 
\beq{415}
 \big(    \widehat{\bf J} +  \rho \widehat{\bf K}\big)\, {\widehat{\bf C}} \,\big(  \rho \widehat{\bf J} +  \rho \widehat{\bf K}\big)
 = 3\kappa \, \widehat{\bf J} + 2\mu \rho^2 \, \widehat{\bf L}+ 2\eta \rho^2 \, \widehat{\bf M}\, . 
\eeq
Thus, $\bar{\lambda}_1=3\kappa $, $\bar{\lambda}_2=\bar{\lambda}_3=\bar{\lambda}_4 = 2\mu \rho^2$,  $\bar{\lambda}_5=\bar{\lambda}_6 = 2\eta\rho^2$, and the eigenvectors are either pure  dilatational ($\alpha_1=1$) or deviatoric ($\alpha_i =0$, $i=2,3,\ldots, 6$).  Therefore, eq. \rf{511} becomes
\beq{513}
\big(3\kappa \big)^{5/6}\, \big(2\mu\big)^{-1/2}\,\big(2\eta\big)^{-1/3}\, \rho^{-5/3} = 1. 
\eeq
Solving for the intermediate variable $\rho$, and evaluating $\mu_R$ and $\kappa_R$ from eqs.
 \rf{214} and \rf{2112} respectively, gives $\kappa_R=\kappa$ and $\mu_R = (\mu^3  \eta^2)^{1/5}$, 
 again in    agreement  with  eq. \rf{25}.

\section{Applications and discussion}

Table 1 lists the computed distance from isotropy of various anisotropic materials, using data from Musgrave \cite{Musgrave}.   Materials of cubic (cub), hexagonal (hex), tetragonal (hex) and orthotropic (ort) symmetry are considered. In each case the moduli of the closest isotropic material were  found using the algorithm described above. The resulting bulk modulus $ \kappa_R$ and  Poisson's ratio $\nu_R$ are tabulated.  

Table 1  ranks the materials  in terms of the Riemannian distance ${\rm d}_R$   of the original anisotropic moduli from the closest isotropic material.  The second column of numbers lists the distance between the closest isotropic materials found using the Riemannian and log-Euclidean norms. That is, 
\beq{091}
{\rm d}_{LR}
\equiv  {\rm d}_{L,R}  \bigr( \widehat{\bf C}_{\rm iso}  (3\kappa_R,2\mu_R),\, \widehat{\bf C}_{\rm iso} (3\kappa_L,2\mu_L)\bigr)
= 
\big[  (\ln \frac{\kappa_L}{\kappa_R})^2 + 5(\ln \frac{\mu_L}{\mu_R})^2 \big]^{1/2}
\, , 
\eeq
which is identically zero for cubic materials. 
The arithmetic  $(\kappa_A,\, \mu_A)$ and harmonic $(\kappa_H,\, \mu_H)$  moduli which minimize the Euclidean distances were also computed, and the Riemannian distance between these two is denoted ${\rm d}_{HA}$.  The  distances ${\rm d}_{RA}$ and ${\rm d}_{RH}$ are the  distances between the closest isotropic material $(\kappa_R,\, \mu_R)$ and the arithmetic and harmonic isotropic approximants, respectively.  All distances listed in Table 1 are  based on the Riemannian metric.  

Note that the distance between the closest materials using ${\rm d}_R$ and ${\rm d}_L$ is less than $0.05$ except for the extremely anisotropic spruce. 
In order to gain some appreciation for the magnitude of the nondimensional distances in Table 1, consider the  distance of any ${\bf P} \in {\cal P}(n)$ from a multiple of itself: 
\beq{092}
{\rm d}_R \big( {\bf P}  ,\, a{\bf P} \big)
= {\rm d}_L \big( {\bf P}  ,\, a{\bf P}\big)
= \sqrt{n}\, \left| \log a \right| , \quad a\in \mathbb{R}_+\, . 
\eeq
Small values of the elastic distance can be identified with  values of $a$ close to unity, specifically 
\beq{093}
a = 1 \pm \tfrac1{\sqrt{6}}{\rm d}_{L,R} + {\rm O}({\rm d}_{L,R}^2) \approx 1 \pm 0.4\, {\rm d}_{L,R}  \, . 
\eeq

Note that the distance ${\rm d}_{HA}$ between the arithmetic and harmonic approximations is generally less than the distance from isotropy ${\rm d}_R$. This is more so for those materials that are closer to isotropy - at the top   of Table 1.  As the material gets further from  isotropy - the lower half of Table 1 - the magnitude of  ${\rm d}_{HA}$ relative to  ${\rm d}_R$ grows as the latter increases.  The two distances are of comparable magnitude for the highly anisotropic materials at the very bottom of the table, such as oak and spruce. 

As a numerical check on the computations, the triangle inequality 
\beq{471}
{\rm d}_{HA}\le {\rm d}_{RA}+ {\rm d}_{RH},
\eeq
was confirmed for each material in Table 1.  Since the three vertices of the triangle are isotropic materials, the  inequality may be written, using \rf{22b}, as
\beq{472}
\big[  (\ln \frac{\kappa_A}{\kappa_H})^2 + 5(\ln \frac{\mu_A}{\mu_H})^2 \big]^{1/2}
\le 
\big[  (\ln \frac{\kappa_A}{\kappa_R})^2 + 5(\ln \frac{\mu_A}{\mu_R})^2 \big]^{1/2}
+
\big[  (\ln \frac{\kappa_R}{\kappa_H})^2 + 5(\ln \frac{\mu_R}{\mu_H})^2 \big]^{1/2}\, . 
\eeq
For cubic materials  $\kappa_A=\kappa_H=\kappa_R$, and consequently  the triangle is a straight line:  
\beq{473}
{\rm d}_{HA}=  {\rm d}_{RA}+ {\rm d}_{RH} \quad \mbox{for cubic materials only}. 
\eeq

  The quantity 
 $({\rm d}_{RA}+ {\rm d}_{RH}-  {\rm d}_{HA})/{\rm d}_{HA} $ was found to be   very small for all the cases considered (and numerically zero for the cubic examples), less than 10$^{-3}$ for all materials considered except barium titanate (1.2$\times$ 10$^{-3}$) and spruce  (2.8$\times$ 10$^{-3}$).  The ``triangle" is almost flat,  indicating that the closest moduli $(\kappa_R,\, \mu_R)$  are in some sense optimally centered  between the arithmetic and harmonic approximations.  Note however, that $\kappa_R$, $\mu_R$  are not equal to the Riemannian mean \cite{Moakher04} of the arithmetic and harmonic approximations, denoted as $\kappa_{AH}$, $\mu_{AH}$.  The Riemannian mean of two elasticity matrices $\widehat{\bf C}_1$ and $\widehat{\bf C}_2$ is $ \widehat{\bf C}_1 (\widehat{\bf C}_1^{-1}\widehat{\bf C}_2)^{1/2}$ \cite{Moakher04}, and consequently the means of the arithmetic and harmonic moduli are  $\kappa_{AH} = (\kappa_{A} \kappa_{H})^{1/2} $, $\mu_{AH}=(\mu_{A} \mu_{H})^{1/2}$.  By considering the case of cubic materials, for which all these quantities  have explicit expressions, it may be shown that $(\mu_{R}-\mu_{AH})(\eta - \mu)> 0$ for $\eta - \mu\ne 0$. 

\section*{Conclusions}
We have presented a method for finding the  isotropic elastic moduli closest to a given material of arbitrary symmetry based on three different metrics.  Unlike the Frobenius (Euclidean) norm, the Riemmanian and log-Euclidean metrics provide unique isotropic moduli. The values obtained according to these two metrics are identical if the comparison medium has cubic symmetry, and are otherwise relatively close. The procedures developed here  for finding the closest isotropic moduli can be generalized to find the closest  material of lower symmetry.  The solution for cubic symmetry  with the cube axes given is presented in the Appendix, and other, lower, symmetries will be considered elsewhere.  Another  generalization of the present problem  is that of  determining the closest  material of cubic or lower symmetry where the symmetry  axes are unrestrained.  These and other challenging questions make this an interesting topic for some time to come.

\appendix  

\Appendix{The closest cubic material}

The cubic stiffness (compliance) closest to $\widehat{\bf C}$ ($\widehat{\bf S}$) in 
 the Euclidean metric ${\rm d}_F$ has moduli $\kappa_A$, $\mu_A$, $\eta_A$ ($\kappa_H$, $\mu_H$ and $\eta_H$), where $\kappa_A$ and $\kappa_H$ are given by \rf{63a} and \rf{64a}, and 
\begin{subequations}\label{67}
\beqa{67a}
   && 6\mu_A   =  \hat{c}_{44} +  \hat{c}_{55} +  \hat{c}_{66} , 
    \qquad 
    6\eta_A   =  \hat{c}_{11}+\hat{c}_{22}+\hat{c}_{33}-\hat{c}_{23}-\hat{c}_{31}-\hat{c}_{12},
   \\
 && 3/({2\mu_H}) =  
  \hat{s}_{44} +  \hat{s}_{55} +  \hat{s}_{66} ,  
  \qquad
  3/({2\eta_H})  = \hat{s}_{11}+\hat{s}_{22}+\hat{s}_{33}-\hat{s}_{23}-\hat{s}_{31}-\hat{s}_{12}   \, .  \quad
 \label{67b}
\eeqa
\end{subequations}

For  the log-Euclidean  norm, we find, using the method for deriving eq. \rf{625}, 
\beq{725}
 \kappa_L  = \tfrac13\, \exp \big( {\rm tr} \, \widehat{\bf J}{\rm Log} ({\widehat{\bf C}}) \big), 
  \quad
\mu_L = \tfrac12\, \exp \big( \tfrac13\, {\rm tr} \, \widehat{\bf L}{\rm Log} ({\widehat{\bf C}}) \big),
 \quad
\eta_L = \tfrac12\, \exp \big( \tfrac12\, {\rm tr} \, \widehat{\bf M}{\rm Log} ({\widehat{\bf C}}) \big). \quad
\eeq
Note the identity, similar to \rf{624}, 
\beq{724}
{\rm det}({\widehat{\bf C}}_{\rm cub} ) = {\rm det}({\widehat{\bf C}})  . 
\eeq

For  the Riemannian norm ${\rm d}_R$ we find that the closest cubic material ${\widehat{\bf C}}_{\rm cub}$ of the form \rf{111b} is determined by three equations: 
 \beq{529}
\prod\limits_{i=1}^n\,   \lambda_i = 1\, ,  \qquad 
\prod\limits_{i=1}^n\,   \lambda_i^{\alpha_i} = 1\, ,  \qquad 
\prod\limits_{i=1}^n\,   \lambda_i^{\beta_i} = 1\, ,  \qquad 
\eeq
where 
\beq{529b}
\alpha_i \equiv {\bf v}_i^T{\widehat{\bf J}}{\bf v}_i,\qquad
\beta_i \equiv {\bf v}_i^T{\widehat{\bf L}}{\bf v}_i,\quad  i=1,2,\ldots n,  
\eeq
and $\{ \lambda_i, \, {\bf v}_i\} $ are the eigenvalues and eigenvectors of
${\widehat{\bf C}}_{\rm cub}^{-1/2}{\widehat{\bf C}} {\widehat{\bf C}}_{\rm cub}^{-1/2}$. The parameters $\alpha_i$ satisfy the same properties as before, including the fact that they sum to unity. 
 Since $\{ {\bf v}_i\}$ form an orthonormal basis, it follows that 
 $\sum_{i=1}^n\,   \beta_i ={\rm dim}\, \widehat{\bf L} = 3 $.  
Furthermore, $\beta_i =0$ if the eigenvector is dilatational.  
The three equations \rf{529} may be reduced to two by assuming the unknown moduli are of the form 
$
{\widehat{\bf C}}_{\rm cub}= 3\kappa_R \,\big(      \widehat{\bf J} + \rho_1^{-2} \widehat{\bf L}
+ \rho_2^{-2} \widehat{\bf M} \big)$.  
Define the modified eigenvalues $ \bar{\lambda}_i = \bar{\lambda}_i (\rho_1,\rho_2)$ to be the 
eigenvalues of 
$\big(      \widehat{\bf J} + \rho_1  \widehat{\bf L}
+ \rho_2  \widehat{\bf M} \big)  \widehat{\bf C}\big(      \widehat{\bf J} + \rho_1  \widehat{\bf L}
+ \rho_2  \widehat{\bf M} \big)$, then  $\kappa_R$ is given by the formula \rf{214}, while 
$\rho_1, \rho_1$ solve the simultaneous equations
\beq{531}
\prod\limits_{i=1}^n\,  \bar{\lambda}_i^{(\alpha_i - 1/n)} = 1,
\qquad
\prod\limits_{i=1}^n\,  \bar{\lambda}_i^{(\beta_i - 1/n)} = 1\, . 
\eeq

\subsection*{Acknowledgments}
It is a pleasure to acknowledge the advice of Maher Moakher.   This paper was motivated by an email discussion initiated by Francis Muir, enlivened by contributions from Sebastien Chevrot, Anthony  Gangi, Klaus Helbig, Albert Tarantola among others. 


\fontsize{8}{9}\selectfont
\vspace{-0.2in}
\begin{table}		\label{t2}
\begin{center}	
\caption{Distance from  isotropy for some materials - data from \cite{Musgrave}. $\kappa_R$ units $10^{10}$ N/m$^2$   }  
\begin{tabular}{lcl cccc rr} \vspace{-0.1in} &&&& &&& &  \\  
\hline  

Material	&	Symm	&	\ d$_R$ \ \  & 100d$_{LR}$ \ \ \ 		&	d$_{RA}$	&	d$_{RH}$ &	d$_{HA}$	&	\ \ \ \  $\nu_R$  \ &	$\kappa_R$
\\ 
\hline 
magnesium	&	hex	&	0.18	&	0.00	&	0.01	&	0.01	&	0.02	&	0.29	&	3.53	\\
diamond	&	cub	&	0.21	&	0	&	0.01	&	0.01	&	0.02	&	0.07	&	44.20	\\
aluminum	&	cub	&	0.21	&	0	&	0.01	&	0.01	&	0.02	&	0.35	&	7.69	\\
beryllium	&	hex	&	0.22	&	0.01	&	0.01	&	0.01	&	0.02	&	0.05	&	11.44	\\
sodium fluoride	&	cub	&	0.29	&	0	&	0.02	&	0.02	&	0.04	&	0.24	&	4.86	\\
ice (H$_2$O) 257$^\circ$K	&	hex	&	0.31	&	0.00	&	0.02	&	0.02	&	0.04	&	0.33	&	0.89	\\
$\beta$-quartz (SiO$_2$)	&	hex	&	0.35	&	0.02	&	0.03	&	0.03	&	0.05	&	0.21	&	5.64	\\
beryllium	&	hex	&	0.37	&	0.23	&	0.03	&	0.03	&	0.06	&	0.26	&	14.41	\\
caesium iodide	&	cub	&	0.37	&	0	&	0.03	&	0.03	&	0.06	&	0.27	&	1.29	\\
sodium chloride	&	cub	&	0.40	&	0	&	0.04	&	0.03	&	0.07	&	0.25	&	2.45	\\
sodium iodide	&	cub	&	0.43	&	0	&	0.04	&	0.04	&	0.08	&	0.25	&	1.46	\\
sodium bromide	&	cub	&	0.44	&	0	&	0.04	&	0.04	&	0.09	&	0.25	&	1.94	\\
caesium bromide	&	cub	&	0.45	&	0	&	0.05	&	0.04	&	0.09	&	0.27	&	1.59	\\
silicon	&	cub	&	0.49	&	0	&	0.05	&	0.05	&	0.11	&	0.22	&	9.78	\\
cobalt	&	hex	&	0.51	&	0.00	&	0.07	&	0.05	&	0.12	&	0.31	&	19.03	\\
silver bromide	&	cub	&	0.52	&	0	&	0.06	&	0.06	&	0.12	&	0.40	&	4.06	\\
germanium	&	cub	&	0.56	&	0	&	0.07	&	0.07	&	0.14	&	0.21	&	7.52	\\
caesium chloride	&	cub	&	0.58	&	0	&	0.08	&	0.07	&	0.15	&	0.27	&	1.83	\\
gallium antimonide	&	cub	&	0.64	&	0	&	0.09	&	0.10	&	0.18	&	0.25	&	5.64	\\
$\alpha$-uranium	&	ort	&	0.68	&	0.37	&	0.10	&	0.10	&	0.20	&	0.20	&	11.28	\\
silver chloride	&	cub	&	0.70	&	0	&	0.11	&	0.10	&	0.22	&	0.41	&	4.42	\\
apatite	&	hex	&	0.72	&	0.11	&	0.10	&	0.13	&	0.22	&	0.21	&	8.43	\\
indium antimonide	&	cub	&	0.75	&	0	&	0.12	&	0.13	&	0.25	&	0.29	&	4.69	\\
potassium fluoride	&	cub	&	0.75	&	0	&	0.13	&	0.12	&	0.25	&	0.28	&	3.19	\\
benzophenone	&	ort	&	0.85	&	1.92	&	0.15	&	0.14	&	0.29	&	0.30	&	5.14	\\
zircon	&	tet	&	0.98	&	0.74	&	0.21	&	0.18	&	0.39	&	0.13	&	1.99	\\
sulphur	&	ort	&	0.98	&	4.13	&	0.20	&	0.18	&	0.39	&	0.34	&	1.88	\\
iron	&	cub	&	0.99	&	0	&	0.20	&	0.23	&	0.43	&	0.30	&	17.05	\\
nickel	&	cub	&	1.01	&	0	&	0.21	&	0.23	&	0.44	&	0.29	&	18.04	\\
cadmium	&	hex	&	1.04	&	3.43	&	0.20	&	0.24	&	0.44	&	0.30	&	5.40	\\
rutile (TiO$_2$)	&	tet	&	1.07	&	0.79	&	0.21	&	0.28	&	0.49	&	0.27	&	21.49	\\
potassium chloride	&	cub	&	1.08	&	0	&	0.27	&	0.24	&	0.50	&	0.28	&	1.78	\\
barium titanate	&	tet	&	1.13	&	3.20	&	0.26	&	0.27	&	0.52	&	0.36	&	17.67	\\
potassium bromide	&	cub	&	1.14	&	0	&	0.30	&	0.26	&	0.56	&	0.29	&	1.58	\\
gold	&	cub	&	1.16	&	0	&	0.27	&	0.31	&	0.58	&	0.42	&	17.28	\\
Rochelle salt	&	ort	&	1.17	&	0.97	&	0.24	&	0.34	&	0.59	&	0.31	&	1.97	\\
zinc	&	hex	&	1.18	&	2.58	&	0.24	&	0.34	&	0.57	&	0.24	&	6.61	\\
white tin	&	tet	&	1.18	&	0.04	&	0.24	&	0.38	&	0.62	&	0.35	&	5.50	\\
ammon. dihyd. phos.	&	tet	&	1.19	&	0.95	&	0.36	&	0.25	&	0.61	&	0.33	&	2.70	\\
silver	&	cub	&	1.21	&	0	&	0.29	&	0.33	&	0.63	&	0.37	&	10.36	\\
potassium iodide	&	cub	&	1.25	&	0	&	0.36	&	0.31	&	0.67	&	0.30	&	1.20	\\
copper	&	cub	&	1.28	&	0	&	0.32	&	0.37	&	0.70	&	0.35	&	13.71	\\
potass. dihyd. phos.	&	tet	&	1.34	&	0.01	&	0.40	&	0.38	&	0.78	&	0.26	&	2.67	\\
$\alpha$-brass	&	cub	&	1.46	&	0	&	0.41	&	0.48	&	0.90	&	0.34	&	11.96	\\
indium 	&	tet	&	1.57	&	0.01	&	0.50	&	0.54	&	1.04	&	0.44	&	4.16	\\
oak	&	ort	&	2.30	&	1.75	&	0.96	&	1.09	&	2.05	&	0.08	&	0.17	\\
$\beta$-brass	&	cub	&	2.34	&	0	&	0.94	&	1.19	&	2.13	&	0.36	&	11.62	\\
spruce	&	ort	&	5.66	&	59.5	&	7.16	&	3.33	&	10.45	&	0.23	&	0.09	\\
\hline 		 
\end{tabular}
\end{center}
\end{table}

\end{document}